\documentclass[sigconf,review=false, nonacm]{acmart}

\usepackage{tabularx}
\usepackage{dcolumn} %Aligning numbers by decimal points in table columns
\newcolumntype{d}[1]{D{.}{.}{#1}}

\usepackage{subcaption}

\usepackage{todonotes}
\let\xtodo\todo
\renewcommand{\todo}[1]{\xtodo[inline,color=green!50]{#1}}

\DeclareRobustCommand{\stepOne}{
  \includegraphics[height=1.2em]{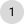}\hspace{.5em}
}

\DeclareRobustCommand{\stepTwo}{
  \includegraphics[height=1.2em]{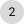}\hspace{.5em}
}

\DeclareRobustCommand{\stepThree}{
  \includegraphics[height=1.2em]{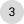}\hspace{.5em}
}

\DeclareRobustCommand{\stepFour}{
  \includegraphics[height=1.2em]{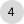}\hspace{.5em}
}

\DeclareRobustCommand{\stepFive}{
  \includegraphics[height=1.2em]{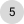}\hspace{.5em}
}

\DeclareRobustCommand{\stepSix}{
  \includegraphics[height=1.2em]{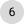}\hspace{.5em}
}
\AtBeginDocument{%
  }

\setcopyright{CC}
\setcctype{by-sa}
\begin{teaserfigure}
   \centering
   \includegraphics[width=.85\linewidth]{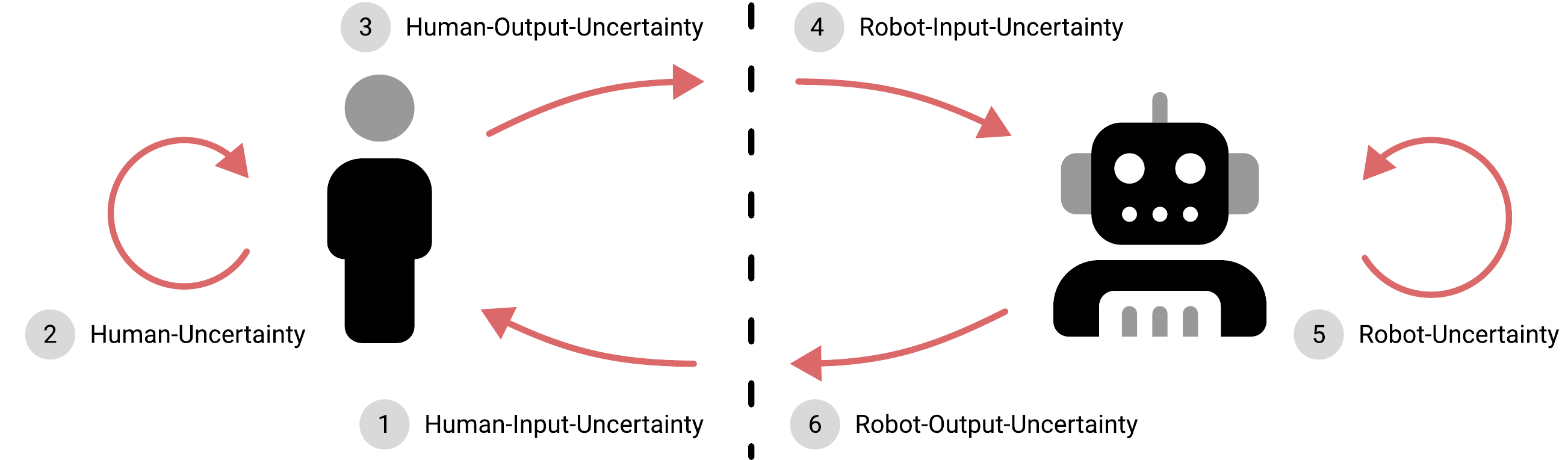}
   \caption{The Human-Robot Uncertainty Loop.}
   \label{fig:uncertaintyloop}
\end{teaserfigure}

\begin{document}

% \title{Challenges and Opportunities for SAR to Understand the User}
% \title{Uncertainty in Human-Robot Interaction: Verbal and Non-Verbal Behavioral Cues }
\title{Understanding the Uncertainty Loop of Human-Robot Interaction}


\settopmatter{authorsperrow=3}

\author{Jan Leusmann}
\orcid{0000-0001-9700-5868}
\affiliation{
  \institution{LMU Munich}
  \city{Munich}
  \postcode{80337}
  \country{Germany}
}
\email{jan.leusmann@ifi.lmu.de}

\author{Chao Wang}
\orcid{0000-0003-1913-2524}
\affiliation{
  \institution{Honda Research Institute Europe}
  \city{Offenbach}
  \country{Germany}
}
\email{chao.wang@honda-ri.de}

\author{Michael Gienger}
\orcid{0000-0001-8036-2519}
\affiliation{
  \institution{Honda Research Institute Europe}
  \city{Offenbach}
  \country{Germany}
}
\email{michael.gienger@honda-ri.de}

\author{Albrecht Schmidt}
\orcid{0000-0003-3890-1990}
\affiliation{%
  \institution{LMU Munich}
  \city{Munich}
  \postcode{80337}
  \country{Germany}}
\email{albrecht.schmidt@ifi.lmu.de}

\author{Sven Mayer}
\orcid{0000-0001-5462-8782}
\affiliation{%
  \institution{LMU Munich}
  \city{Munich}
  \postcode{80337}
  \country{Germany}}
\email{info@sven-mayer.com}

\renewcommand{\shortauthors}{Leusmann et al.}

\begin{abstract}
Recently the field of Human-Robot Interaction gained popularity due to the wide range of possibilities of how robots can support humans during daily tasks. One such form is supportive robots, socially assistive robots built explicitly for communicating with humans, e.g., as service robots or personal companions. As they understand humans through artificial intelligence, these robots can sometimes make wrong assumptions about the humans' current state and give an unexpected response. In human-human conversations, unexpected responses happen frequently. However, it is currently unclear how such robots should act if they understand that human did not expect their response or even show the uncertainty of their response in the first place. For this, we explore the different forms of potential uncertainties during human-robot conversations and how humanoids can communicate these uncertainties through verbal and non-verbal cues.
\end{abstract}




\maketitle

\section{Introduction}
Robots are becoming a more and more ubiquitous part of our lives. While already being used in industry to perform repetitive tasks or tasks that humans are not capable of, they are recently evolving into increasingly sophisticated and integrated systems. Due to the emergence of artificial intelligence (AI) technologies, robots can perform increasingly complex tasks with the ability to react to observed environmental actions. In the future domestic robots will actively support users in their homes with various tasks. 

Currently, the field of social assistive robots (SARs) is developing rapidly. These robots are mostly designed to provide support and assistance to humans in fields like healthcare~\cite{cooper2021social}, education~\cite{PAPADOPOULOS2020103924}, and social interactions~\cite{feil2009socially}. In all these contexts, the SARs are able to detect users' questions or actions and react in a certain way to them. However, as these reactions are model-based, the detected user input classification accuracy can be low or even wrong. Hence, the interactions from the SARs usually happen with measurable certainty. However, errors can also happen even though the accuracy of the predicted action is high, e.g., when the users themselves give miss-interpretable prompts. Uncertainty happens regularly in human-human interaction. Humans can express this uncertainty easily through verbal and non-verbal communication cues~\cite{mavridis2015review}. Thus, for fluid interaction with these robots, they must be able to deal with uncertainty from both the users and themselves. On the one hand, it is crucial that users can interpret verbal and non-verbal interaction cues correctly and map them toward the certainty of the robots. On the other hand, robots need to be able to interpret the users' uncertainty, understand uncertainty, and also to be able to express uncertainty.

In this work, we explore and compare the various kinds of uncertainty in human-robot interaction (HRI). Previous work found that robot uncertainty should be communicated transparently to users~\cite{moon2010using}. Various machine learning (ML) models are able to classify human uncertainty~\cite{cumbal2020Uncertainty, grubov2022frequencyspace, duarte2020human, hramov2018artificial}. As there are different possible uncertainties during human-robot interaction, we first look at each step of the interaction loop in detail. We then discuss the advantages and disadvantages of these cues being human-like.

\section{The Human-Robot Uncertainty Loop}
\label{sec:loop}
We propose that this interaction concept can be envisioned as an uncertainty loop, see \autoref{fig:uncertaintyloop}. In the following, we state our explanations and findings for each of the six steps of this loop. 
% As we can not influence the human, we focus on the investigation of the robot side, i.e., steps 4, 5, and 6 but draw conclusions from human reasoning.

\subsection*{\stepOne Human-Input-Uncertainty}
\textit{The ability of humans to understand the uncertainty expressed by the robot.} The mental model of humans of a robot's understanding of the environment is generally connected to their own knowledge~\cite{lee2005human}. As collaborative robots are still a relatively new research topic, most humans' trust, and therefore, certainty or understanding of robots is still low~\cite{washburn2020robot}. Thus, if users do not understand the actions performed by a robot, they cannot reliably predict the outcome of this action. Humanoid robots mimicking human cues in human-human interaction can increase this understanding through, e.g., eye contact or showing engagement when handing over an object~\cite{grigore2013joint}. Non-humanoid robots can also try to mimic human-like movements. Robots can also show their intent through different output channels~\cite{chadalavada2015that, augustsson2014how}. \citet{dragan2013legibility} proposed that robot motion can be made legible and, thus, predictable, which is a crucial part of seamless collaborative human-robot interaction.

\subsection*{\stepTwo Human-Uncertainty}
\textit{Humans' intrinsic uncertainty about their own actions. } 
Humans are naturally prone to make mistakes, unpredictable decisions, or change their minds~\cite{bland2012different}. When sure about something, confidence in one's own decisions is higher~\cite{navajas2017idiosyncratic}. However, especially in unknown scenarios, humans' intrinsical uncertainty is higher~\cite{milburn1977decision} due to the lack of information. In our HRI scenario, most humans still need to learn how to interact with robots correctly and how to understand their actions. 

\subsection*{\stepThree Human-Output-Uncertainty}
\textit{The ability of humans to express uncertainty.} Humans use verbal and non-verbal cues to express their level of certainty, both when speaking and body language when performing an action. Often we use prosodic cues, like filler words, to express uncertainty. \citet{brennan1995feeling} showed that others could make reasonable estimations about the uncertainty level of the speaker. However, verbal uncertainty cues are overshadowed by non-verbal cues~\cite{borras-comes2011perceiving}. For example, non-verbal cues to show uncertainty include shrugging, frowning, hesitating, or palm-up gestures~\cite{givens2006nonverbal}. 

\subsection*{\stepFour Robot-Input-Uncertainty}
\textit{The ability of the robot to understand the uncertainty expressed by the human.} With this, we describe how AI systems can measure the uncertainty level of users. Literature suggests using facial action units~\cite{cumbal2020Uncertainty}, EEG analysis during visual decision-making~\cite{grubov2022frequencyspace}, detection of non-verbal gestures~\cite{duarte2020human}, or brain states~\cite{hramov2018artificial} to detect human uncertainty. If the robot, especially a SAR, detects high uncertainty in a user prompt, it can react in certain ways by asking questions to make the user more certain or by giving reassuring cues. For reinforcement learning approaches, we can also use the level of uncertainty as a metric for the agent~\cite{scherf2022learning}. 
In general, detecting the uncertainty levels of humans is possible. 
We noted that as uncertainty detection is also done with several ML models, they also have internal uncertainty. Humans are able to react to the expression of the uncertainties of others. Robots should also be able to react to and express uncertainty to enable fluid human-robot interaction.
However, we do not perceive robots as other humans. Thus, we need to determine whether robots should express a deep knowledge about the users' uncertainty level. 

\subsection*{\stepFive Robot-Uncertainty}
\textit{The uncertainty of the robot to understand the world accurately. }
This describes the robots' or AI systems' own level of uncertainty. In contrast to human uncertainty, this can be directly ``measured,'' as it is just the prediction accuracy of an action. Earlier, this was done with essential machine learning approaches \cite{albrecht1997bayesian, bui2002policy, yu2015fast}. Today, mainly deep neural networks are used \cite{xu2020gtad, zhang2019comprehensive}. Moreover, although today's models are already way more accurate at classifying human actions, models are still prone to misclassifications. They also do not know every single human action but are usually trained on a specific subset of actions. \citet{trick2019multimodal} combined classification results from different input data sources (speech, gesture, gaze, and objects) to reduce uncertainty in an intention recognition task for collaborative assistive robots. Humans gain a world understanding through experiences. Reinforcement learning can resemble this for AI agents.

\subsection*{ \stepSix Robot-Output-Uncertainty}
\textit{The ability of the robot to express uncertainty. }
Same as human-output-uncertainty, this describes how the robot can express the level of uncertainty of their upcoming interaction. Again this interaction can be either verbal or non-verbal. However, the cues could mimic human behavior or be totally artificial. For example, we can use facial expressions to convey emotional states; thus, we can also use them to show uncertainty. Gestures or movements, in general, can also be used to convey certainty. For example, \citet{hough2017It} mapped the confidence of robot decision onto its movement speed. They found that users are able to understand the level of certainty of the robot. We can also use the tone and pitch of artificially generated voices of robots to convey uncertainty~\cite{szekely2017Synthesising}. However, robots can also come in non-humanoid shapes, e.g., a robotic arm can assist with kitchen tasks \cite{oechsner2022challenges}. Here, we can leverage artificially created gestures or output modalities like light to convey goals or uncertainty~\cite{chadalavada2015that, augustsson2014how}. For instance, \citet{wang2023explainable} used an augmented reality representation to indicate future actions.

\section{Discussion}
In this work, we investigate uncertainty in the context of HRI. In this context, both humans and robots can be uncertain about their decisions. Humans are often intrinsically unsure about their decisions and, thus, uncertain. Robots, or AI systems in general, base their decision on classification accuracies - which can range from low to high values. We can neither influence human uncertainty levels nor assure that classification models are always right, especially in classifying human behavior, where there is underlying uncertainty. With our proposed human-robot uncertainty loop (see \autoref{fig:uncertaintyloop}), however, we can look at various parts of this process in greater detail and better understand how we can influence the robots' behavior to improve interaction when either the human or robot has a high level of uncertainty. 

In human-human interaction, uncertainty is often expressed and also understood subconsciously. Using prosodic cues is not done actively, but using filler words typically indicates higher uncertainty, e.g., in a presentation. Additionally, not only as a speaker but also as a listener, we usually only notice the filler words once someone makes us aware of them. With Duplex, Google developed a natural conversation agent, which among other things, used filler words to sound more natural\footnote{\url{https://ai.googleblog.com/2018/05/duplex-ai-system-for-natural-conversation.html}}. However, \citet{mori2012Uncanny} argued that this level of an AI agent being able to act very human-like was in the uncanny valley. \citet{thepsoonthorn2021Exploration} propose similar findings for non-verbal behavioral cues. 

In general, it would be obvious to let humanoid robots express uncertainty similar to humans. However, research suggests it might be unsuitable in at least some cases~\cite{fox2021Relationship, giger2019humanization}. In other cases, it is impossible to use human-like cues, especially if the robot is not humanoid. Nevertheless, even in these cases, we are good at understanding non-verbal cues and translating them towards certain emotional states, e.g., understanding R2-D2 or BB-8 from Star Wars, even though their non-verbal cues are very simplistic. Together with these findings, we propose investigating different kinds of verbal and non-verbal cues for steps 4, 5, and 6 from \autoref{sec:loop}. Here, we need to consider that by mimicking humans too much; users will perceive robots quickly as uncanny. 

% Different other factors can also influence humans' uncertainty during interaction with robots, e.g. properties of objects during handover~\cite{}, example2, example3. 
% \todo{Perhaps would be interesting to expand upon the point “if robots should express a deep knowledge about the users’ uncertainty level” with regard to the balance of this with the potential increase in miscommunications and errors in interaction.}

\section{Summary}
In this work, we propose the human-robot uncertainty loop to look at uncertainty during this interaction in a more fine granular way. Both humans and robots can be uncertain about their world understanding and correctly understanding their conversation partner. We found that robots have two options for giving behavioral cues regarding uncertainty. First, reacting to humans' uncertainty and second, conveying their own uncertainty. Both of these can be human-like or artificially generated. On the one hand, we need to investigate which, if, and how humans understand these cues conveying uncertainty. On the other hand, we also need to measure whether users find human-like cues uncanny, as previous research suggests. 
By understanding uncertainty better, especially when and how to react to human and robot uncertainty, with verbal and non-verbal cues, we can ensure a more explainable and understandable interaction.

\bibliographystyle{ACM-Reference-Format}
\bibliography{bibliography}

\end{document}